\documentclass[epsfig]{mn2e}

\input{psfig.sty}
\usepackage{color}
\usepackage{graphicx}

\title[Mg,Ca,N,C/Fe ratios in elliptical galaxies]
{Stellar mass-loss, rotation and the chemical enrichment of early type galaxies}

\author[A. Pipino et al]{Antonio Pipino$^{1,2}$, 
Cristina Chiappini$^{3,4}$, Genevieve Graves$^5$ and Francesca Matteucci$^{6,4}$\\
$^1$Physics \& Astronomy, University of Southern California, Los Angeles 90089-0484, USA\\
$^2$Astrophysics, University of Oxford, Denys Wilkinson Building,
    Keble Road, Oxford, OX1 3RH, U.K.\\
$^3$Observatoire de Gen\`eve, Universit\'e de Gen\`eve,
51 Chemin de Mailletes, CH1290, Sauverny, Switzerland\\
$^4$INAF--Osservatorio Astronomico di Trieste,
    Via G.B. Tiepolo, 11, I-34127, Trieste, Italy\\
$^5$UCO/Lick Observatory, University of California, Santa Cruz, CA 95064.\\
$^6$Dipartimento di Astronomia, Universita di Trieste,
    Via G.B. Tiepolo, 11, I-34127, Trieste, Italy}
\date{Accepted 2009 March 25.  Received 2009 March 13; in original form 2008 December 20}

\begin{document}
\maketitle

\begin{abstract}
We present a comparison between the [Ca,C,N/Fe]-mass relations
observed in local spheroids and the results of a chemical evolution
model which already successfully reproduces the [Mg/Fe]-mass and the
[Fe/H]-mass relations in these systems.  We find that the [Ca/Fe]-mass
relation is naturally explained by such a model without any additional
assumption.  In particular, the observed under-abundance of Ca with
respect to Mg can be attributed to the different contributions from
supernovae Type Ia and supernovae Type II to the nucleosynthesis of
these two elements.  For C and N, we consider new stellar yields that
take into account stellar mass loss and rotation.  These yields have
been shown to successfully reproduce the C and N abundances in Milky
Way { metal-poor} stars.  The use of these new stellar yields produces a  good
agreement between the chemical evolution model predictions and the
integrated stellar population observations for C.  In the case of N,
the inclusion of fast rotators and stellar mass-loss nucleosynthesis
prescriptions improves our predictions for the slope of the [N/Fe]
vs. $\sigma$ relation, but a zero point discrepancy of 0.3~dex
remains.  This discrepancy cannot be removed, either by increasing the
N yields or by assuming a larger amount of fast rotators in spheroids,
because in both cases this leads to an overproduction of the N
abundances in the gas phase in these galaxies at high redshift
(e.g. the Lyman Break Galaxy MS1512 cB-58).  This work demonstrates
that current stellar yields are unable to simultaneously reproduce the
large mean stellar [$\langle$N/Fe$\rangle$] ratios inferred from
integrated spectra of elliptical galaxies in SDSS and the low N abundance
measured in the gas of high redshift spheroids from absorption lines.
 However, since { chemical evolution models for the Milky Way computed with the Geneva stellar yields constitute at present the only way to account for the N/O, C/O and $^{12}$C/$^{13}$C abundance ratios observed in very metal-poor halo stars}
, it seems reasonable to suggest that there may be uncertainties in either the inferred stellar or gas-phase N abundances at the level of $\sim$0.3~dex.  
\end{abstract}


\section{Introduction}

Abundance ratios can be used to constrain both galaxy formation
scenarios and stellar nucleosynthesis. So far, in studies of
elliptical galaxies, abundance ratios have been used with the former
aim. Indeed, increasing evidence has accumulated over the past decade
that the [Mg/Fe] ratio is super-solar in the cores of bright galaxies
and increases with galactic mass (e.g. Faber et al., 1992, Worthey et
al. 1992).  It is also well established that elliptical galaxies obey
the mass-metallicity relation, namely the stars of the brighter
spheroids are more metal rich than those of the less luminous ones (Carollo et
al., 1993).  These two relations, along with the evolution of the
luminosity function (e.g.  Bundy et al. 2006) led to the so-called
downsizing picture: more massive galaxies form faster and earlier than
less massive ones (Matteucci 1994).

Unfortunately, due to the fact the we cannot resolve stars in
elliptical galaxies and because of the uncertainties in modelling
line-strength indices as a function of abundance ratios, elliptical
galaxies have not yet been used to constrain stellar nucleosynthesis
calculations. Up to now, such studies have been restricted to the
Milky Way (MW; e.g. Chiappini et al. 2003a,b, Cescutti et al. 2006)
and its satellites (e.g. Lanfranchi et al. 2006, 2008 for dwarf
spheroidals; Pompeia et al. 2008 for the Large Magellanic Cloud), and
Damped Lyman Alpha systems (e.g. Dessauges-Zavadsky et al. 2007).

As an example, metal poor halo stars in the MW are observed to have
high levels of N/O and C/O (Spite et al. 2005, Akerman et
al. 2004, { Fabbian et al. 2009}). The observed trends in C/O and N/O versus O/H at very low
metallicities cannot be reproduced by chemical evolution models in
which standard yields are adopted (see Chiappini et al. 2005 for a
detailed discussion).  In fact, due to the very low metallicities of
such stars, the high levels of N/O would have to be the result of
nucleosynthesis in metal-poor massive stars, suggesting these objects
must produce primary nitrogen in non-negligible quantities\footnote{At
  least a factor of a few hundred more than what was already predicted
  by the models of Meynet \& Maeder (2002) with rotational velocities
  of 300 km s$^{-1}$ (see Chiappini et al. 2005 for details).}. It has
recently been shown that fast stellar rotation is a promising
mechanism for producing primary nitrogen in metal-poor massive stars
(e.g. Meynet et al. 2006, Hirschi 2007).  Chemical evolution models
which include the yields of fast rotating models at low metallicity
($Z=10^{-8}$ { by mass}) can account for the observations in normal
metal-poor halo stars (Chiappini et al. 2006a,b, 2008).

In principle, there is no reason why such fast rotators should occur
only in the MW halo. These stars could also have left their imprints
in DLAs and sub-DLAs (the most metal poor DLAs found at present). In
fact, Pettini et al. (2008) find that these metal-poor high-redshift
systems also show large C/O and N/O ratios, compatible with what is
found in the MW halo.  Given these successes, it is important to
determine whether the stellar yields adopted in Chiappini et
al. (2006a) can also give consistent results for { other systems
whose star formation histories differ from that of the solar vicinity and the MW,
such as DLAs} (Chiappini et al. in prep) and elliptical galaxies (this
work).

The large number of galaxies observed in the recent Sloan Digital Sky
Survey (SDSS; York et~al. 2000) make it possible to construct very
high S/N spectra for typical elliptical galaxies by stacking together
the spectra of many similar objects.  The resulting data are of high
enough quality to {enable the } study { of } line-strength ratios for multiple different
elements on a more robust basis (Graves et al. 2007).  At the same
time, Schiavon (2007) has produced a new class of stellar population
models which take into account the effect of various
abundance patterns.  These models allow us to estimate not only the
total metallicity, age, and [$\alpha$/Fe] of galaxies, but also the
relative abundances of several other elements, such as carbon,
nitrogen and calcium.

Graves et al. (2007) find that many properties of the stellar
populations in red sequence galaxies (such as age,
[$\langle$Fe/H$\rangle$] and [$\langle$Mg/Fe$\rangle$]) vary with
galaxy velocity dispersion ($\sigma$). In particular, more massive
galaxies are typically older, more metal-rich, and more enhanced in
Mg, as predicted by the chemical evolution models of Matteucci (1994)
and in agreement with previous work (e.g., Trager et~al. 2000b,
Bernardi et~al. 2003, Nelan et~al. 2005).  Using the newly-developed
capabilities of the Schiavon (2007) models, they { further} measure
[$\langle$Ca/Fe$\rangle$], [$\langle$C/Fe$\rangle$], and
[$\langle$N/Fe$\rangle$].  They find that Ca scales with Fe and, more
surprising, that [$\langle$C/Fe$\rangle$] and [$\langle$N/Fe$\rangle$]
increase with $\sigma$ more strongly than 
[$\langle$Mg/Fe$\rangle$] does.

These results are crucial new constraints to study the role played by stars
of different masses in the chemical enrichment of elliptical galaxies.
This investigation aims at reproducing and understanding these abundance
trends by means of a self-consistent chemical evolution model, whose
predictions already match many previous observations.  In this way we
can, for the first time, use several
 abundance ratios measured in the
stars of elliptical galaxies as constraints on nucleosynthesis
calculations.

This paper is organized as follows.  We first review the observational
data and their uncertainties in Section 2, then present the main
ingredients of our chemical evolution models in Section 3.  In Section
4, we present our results and then summarize our main conclusions in
Section 5.

\section{Observations}

\subsection{The Data}

Graves et~al. (2007) use a sample of $\sim$6,000 galaxies from the
SDSS Main Galaxy Survey (Strauss et~al. 2002) to determine the typical
stellar population properties of elliptical galaxies as a function of
$\sigma$.  Their sample compares the stellar populations of quiescent
elliptical galaxies (those with no emission lines in their spectra) to
those of elliptical galaxies that host emission from low ionization
nuclear emission-line regions (LINERs).  In this work, we are only
{ interested in} quiescent galaxies.  Graves et~al. sort galaxies
into six $\sigma$ bins, then stack the spectra of the galaxies in each
bin to obtain very high S/N mean spectra.  They then measure the
strength of multiple absorption features in the spectra (using the
Lick/IDS index system pioneered by Burstein et~al. 1984) and use the
models of Schiavon (2007) to convert the line strengths into measured
values of mean, luminosity-weighted age, [$\langle$Fe/H$\rangle$],
[$\langle$Mg/Fe$\rangle$], [$\langle$Ca/Fe$\rangle$],
[$\langle$C/Fe$\rangle$], and [$\langle$N/Fe$\rangle$] for the stellar
populations.  

In agreement with many previous authors, Graves et~al. find that age,
[$\langle$Fe/H$\rangle$], and [$\langle$Mg/Fe$\rangle$] all increase
with $\sigma$ (see their Table 8 for a detailed comparison with
previous results).  In particular, they find that
[$\langle$Mg/Fe$\rangle$] ranges from $+0.13$ to $+0.26$ as $\sigma$
goes from 100 km s$^{-1}$ to 250 km s$^{-1}$.  They furthermore find
that [$\langle$Ca/Fe$\rangle$] has only very weak dependence on
$\sigma$ and is nearly solar for all galaxies, with values in the
range $-0.01 < $ [$\langle$Ca/Fe$\rangle$] $ < +0.03$ for galaxies with
100 km s$^{-1} < \sigma < 250$ km s$^{-1}$.

Some of the most interesting results from the Graves et~al. study, and
the ones which motivated this analysis, were the observed trends in
[$\langle$C/Fe$\rangle$] and [$\langle$N/Fe$\rangle$].  Both of these
elements show strong enhancement with increasing $\sigma$ (see Figures
1 and 2).

An overabundance of C and N in more massive ellipticals has been
suggested by previous authors, but strong evidence has been lacking
and there has been considerable confusion in the literature.  The
first hints came from Trager et al. (2000a)'s analysis of a sample of
local ellipticals. They suggested that, in order to best fit their
absorption line measurements for elliptical galaxies, C should be
enhanced along with Mg and possibly O.  Trager et~al. therefore
included C (as well as N) in the group of ``enhanced'' elements when
constructing their stellar population models but did not attempt to
{quantitatively} measure [$\langle$C/Fe$\rangle$] or
[$\langle$N/Fe$\rangle$].  Later, Sanchez-Blazquez et al (2003) found
a positive correlation between the strength of a CN absorption band
and $\sigma$, which they interpreted as evidence for further
enhancement in C and N with increasing $\sigma$.  They did not attempt
a quantitative conversion into abundance ratios.

Sanchez-Blazquez et~al. further presented some evidence that galaxies
in denser environments were somewhat less enriched in C and N with
respect to their low density counterparts.  However, Toloba
et~al. (2009) have recently claimed that this correlation disappears
if one uses an NH feature at shorter wavelength.  To further
complicate matters, Clemens et~al. (2006) corroborate the
environment-correlation proposed by Sanchez-Blazquez et~al., but find
an {\it anti-correlation} of [$\langle$C/H$\rangle$] with galactic
mass, in conflict with the other studies reported here.

The Graves et~al. (2007) results are based on a large, homogeneous data set
and take advantage of new advances in stellar population modelling.
Furthermore, { Graves \& Schiavon (2008)} have tested their method on integrated spectra of MW
globular clusters.  For a small number of
clusters, [C/Fe] and [N/Fe] have been measured in spectra of
individual cluster stars. { The latter authors} demonstrate that
the values of [$\langle$C/Fe$\rangle$] and [$\langle$N/Fe$\rangle$]
measured in integrated cluster spectra using their method match the
average abundance determinations from individual cluster stars to
within { $\pm0.04$ dex}.  We therefore consider their abundance
measurements to be reliable.  However, the reader should bear in mind
that there is not yet a consensus on C and N abundances in elliptical
galaxies.

\subsection{Uncertainties in deriving [C/Fe] and [N/Fe] from integrated spectra}

The key results of this analysis consist of a comparison of the
observed values of [$\langle$C/Fe$\rangle$] and
[$\langle$N/Fe$\rangle$] from Graves et~al. (2007) versus those obtain
from chemical evolution models.  In this context, it is important to
review the uncertainties in the derivation of [$\langle$C/Fe$\rangle$]
and [$\langle$N/Fe$\rangle$] using the method of Graves \& Schiavon
(2008).  We list these in order of their likely importance, from most
to least important.

\begin{itemize}

\item[i)] {\it Modelling [$\langle$N/Fe$\rangle$] using a CN molecular absorption
  feature.} The abundance modelling process used in Graves et
  al. (2007) uses a CN molecular absorption feature to determine
  [$\langle$N/Fe$\rangle$], after [$\langle$C/Fe$\rangle$] has been fixed based on a C$_2$
  feature (see Graves \& Schiavon 2008 for details).  This means that
  the measured value of [$\langle$N/Fe$\rangle$] depends on the accuracy of the
  [$\langle$C/Fe$\rangle$] determination and therefore has a larger associated
  uncertainty than the other abundance ratio measurements.  In Graves
  \& Schiavon (2008), this method is tested in globular clusters and
  is shown to reproduce the average [$\langle$N/Fe$\rangle$] determined from
  individual cluster stars to within 0.04 dex.  This suggests that the
  modelling process is accurate, however the test has only been
  performed on two globular clusters to date. Recently, Toloba et
  al. (2009) suggested that the N abundance does not strongly
  correlate with mass by using a NH feature in the near UV, which may
  be a cleaner discriminant of N abundance.  However, Toloba et
  al. did not attempt any conversion between the index and the actual
  N abundance, therefore we cannot use their data as a further
  constraint for our models.

\item[ii)] {\it Uncertainty in the [O/Fe] abundance.}  The
  [$\langle$C/Fe$\rangle$] and [$\langle$N/Fe$\rangle$] stellar
  abundance ratios are measured using the C$_2$4668 and CN$_1$
  molecular absorption features.  Both of lines are sensitive to the
  abundance of O, which competes with the formation of C$_2$ and CN.
  Thus higher O abundances will inhibit the formation of these
  molecules, resulting in weaker absorption features for fixed values
  of [C/Fe] and [N/Fe].  Unfortunately, O is currently unmeasurable in
  spectra of unresolved stellar populations.  The abundances presented
  here are computed assuming solar { O/Fe} abundance { ratio} (i.e., [O/Fe] = 0.0).
  However, early type galaxies are known to be enhanced in
  $\alpha$-elements and likely have super-solar [O/Fe].  Accounting
  for the likely super-solar [O/Fe] will have the effect of {\it
    increasing} the measured stellar values of
  [$\langle$C/Fe$\rangle$] and [$\langle$N/Fe$\rangle$], since larger
  C and N abundances will be needed to produce the same absorption
  line strengths.  As will be shown, the measured values of [C/Fe] and
  [N/Fe] are relatively high already.  Also, recent evidence from MW bulge
  stars suggests that [O/Fe] may be $\sim$solar for metal-rich
  bulge stars, despite their super-solar values of [Mg/Fe] (Fulbright
  et~al. 2007).

\item[iii)] {\it The effects of rotation on evolutionary tracks.}
  There is some indication that stars with fast rotation have bluer
  evolutionary tracks than non-rotating stars of the same mass
  (Meynet, private communication), which will affect the stellar
  population analysis of these galaxies. However, while the
  temperature differences should have a substantial impact on age
  measurements, they should have little effect on the abundance ratios.
  This can be seen in the comparison of solar-scale and
  $\alpha$-enhanced evolutionary tracks in Schiavon (2007) and Graves
  et al. (2007), where turn-off temperatures and age measurements are
  strongly affected but abundance ratios are not.

\item[iv)] {\it The effects of emission infill.}  The galaxies in
Graves et al. (2007) have been selected to have no detectable emission
lines in their spectra, based on the strong emission lines at
H$\alpha$ and [OII]$\lambda$3727.  Thus the effects of
H$\delta$ emission are unlikely to substantially contaminate the CN
absorption index in these galaxies.  For galaxies in Graves et
al. (2007) that did contain significant emission (not used in this
work), variations in the emission infill correction resulted in
differences of $< 0.01$ dex in [$\langle$N/Fe$\rangle$]. 

\item[v)] {\it Lick index zero-point offsets.} The galaxy data from
  Graves et al. (2007) are not zero-point shifted to the original
  Lick/IDS system.  However, this should not have a significant effect
  on the abundance determinations.  Both the SDSS data and the
  Schiavon (2007) stellar population models used to interpret them are
  based on flux-calibrated spectra.  Schiavon (2007) demonstrates that
  the zero-point offsets between Lick indices measured in
  flux-calibrated spectra are extremely small (see Figure 2 of that
  work), with the exception of the indices Fe5015, Fe4383,
  H$\gamma_A$, and H$\delta_A$, none of which are used the in
  abundance analysis of Graves et al.

\end{itemize}

\section{The model}

\subsection{The PM04 chemical evolution model}

The adopted chemical evolution model is an updated version of the
multi-zone model of Pipino \& Matteucci (2004, PM04 hereafter).  We
calculate the evolution of element abundances by means of the equation
of chemical evolution (see e.g. Matteucci $\&$ Greggio 1986 for a
comprehensive discussion of this equation).  The Initial Mass Function
(IMF) $\phi (m)\propto m^{-(1+x)}$ is normalized to unity in the mass
interval $0.1 -100 M_{\odot}$.  We adopt an IMF with $x=1.35$
(Salpeter, 1955).

We adopt the following law for the star formation rate:
\begin{equation}
\psi (t)= \nu \cdot M_{gas} (t)\, ,
\end{equation} 
where the star formation rate $\psi (t)$ is assumed to be proportional
to the gas mass via a constant $\nu$ which represents the star
formation efficiency.  { We assume $\nu = \nu_{PM04}$, i.e., as in
  model II of PM04 where $\nu$ is an increasing function of the
  galactic mass.

The star formation history of a model galaxy is determined by the
interplay between the infall time-scale, the star formation efficiency
and the onset of the galactic wind (i.e. the energetic feedback from
SNe and stellar winds).  We assume that $\psi=0$ after the development
of the galactic wind.

A fundamental component of this model is the detailed calculation of
supernova (SN) explosion rates.  For Type Ia SNe, we assume a
progenitor model made of a C-O white dwarf plus a red giant (Greggio
$\&$ Renzini, 1983; Matteucci $\&$ Greggio, 1986).  The predicted Type
Ia SN explosion rate is constrained to reproduce the present day
observed value (Mannucci et al., 2008)

Here we adopt the same formulation for the feedback as in Pipino et
al. (2002), to which we refer the reader. In brief, we consider a
$\sim$ 20\% mean efficiency in energy transfer from the SN into the
interstellar medium.  In addition, we define the onset of the galactic
wind} ($t_{gw}$) as the time at which the energy input by SNe exceeds
the gas binding energy (for details see PM04 and Pipino et al. 2002).
The wind carries out the residual gas from the galaxies, thus
inhibiting further star formation.

In PM04, we simulated the creation of the spheroid as due to the
collapse of either a single large gas cloud or several smaller gas
lumps.  The inclusion of gas infall makes the star formation rate
start at a lower value than in the closed box case, reach a maximum,
and then decrease as the gas is used up.

The infall term gives the rate at which primordial gas is accreted by the proto-galaxy.
The adopted expression is:
\begin{equation}
({d G_i (t) \over d t})_{infall}= X_{i,infall} C e^{-{t \over \tau}}\, ,
\end{equation}
where { $G_i (t)$ is the mass density of element \emph{i} at time
\emph{t}} and $X_{i,infall}$ describes the chemical composition of the
accreted gas, which is assumed to be primordial.  $C$ is a constant
obtained by integrating the infall law over time and requiring that
$\sim 90\%$ of the initial gas has been accreted at $t_{gw}$ (in fact,
we halt the infall of the gas at the onset of the galactic wind).
Finally, $\tau$ is the infall time-scale.

In order to compare our predicted abundances with the observed ones,
we must compute the mean stellar abundance of the element X
($\langle$X/H$\rangle \equiv \langle Z_X \rangle$), defined as (Pagel
$\&$ Patchett 1975):
\begin{equation}
\langle Z_X \rangle ={1\over S_0} \int_0^{S_0} Z_X (S) dS\, ,
\label{pp75}
\end{equation}
where $S_0$ is the total mass of stars ever born contributing to the
light at z=0. For massive ellipticals, results obtained by averaging
over the stellar mass are very close to those obtained by averaging
over the stellar luminosity at z=0 (the difference for the models
presented in this paper is typically less that 0.05 dex), since the
indices are weighted by V-band luminosity (see e.g. Matteucci et al.,
1998).

\subsection{General overview of the models}

We run models for elliptical galaxies with masses $10^{10}M_{\odot}$
and $10^{12}M_{\odot}$.  The basic features common to all models are
listed in Table 1, where the input luminous mass, effective radius,
star formation efficiency, infall time-scale and time of the galactic
wind onset are listed in columns 1-5, respectively.  $M_{lum}$ is the
\emph{nominal} mass of the object, i.e. the mass of the initial gas
cloud, with the infall law normalized such that 90\% of the mass is
accreted between $t = 0$ and $t \sim t_{gw}$. The mass in stars at
$z=0$ is $\sim 0.2-0.4$ $M_{lum}$ for all models and the velocity
dispersion $\sigma$ is evaluated from the relation $M=4.65\cdot 10^5\,
(\sigma/$km s$^{-1})^2\, (R_{eff}/$kpc$)\, M_{\odot}$ (Burstein et
al., 1997).  The effective radius $R_{eff}$ is the final one, achieved
when the collapse is over.

Parameters such as the star formation efficiency ($\nu$), the infall
time-scale ($\tau$), the IMF, and the fraction of binary systems that
give rise to a SNIa explosion have been taken to be the same as in the
best model of PM04. { The latter authors} show that these choices lead to good
agreement with a large set of optical observables.  In the present
work, we can study for the first time the chemical evolution in much
finer detail.  The modifications to the standard PM04 nucleosynthesis
presented here are modest adjustments affecting Ca, C, and N
abundances.  They do not alter the satisfactory agreement between the
observations discussed in PM04 and their best model predictions.

\subsection{Stellar Yields}

Although in the present work we will discuss only Mg, Fe, Ca, N and C,
our code follows in detail the evolution of 21 chemical elements, for
which we need to adopt detailed stellar nucleosynthesis prescriptions.

We first define a \emph{fiducial} model (Model PM04) with the same
nucleosynthesis prescriptions as in PM04, namely:

\begin{enumerate}
\item
For single low and intermediate mass stars ($0.8 \le M/M_{\odot} \le
8$) we make use of the yields of van den Hoek $\&$ Groenewegen (1997,
vdHG) as a function of metallicity, with the mass loss parameter
$\eta_{AGB}$ dependent on the metallicity as follows: $\eta_{\rm
  AGB}=1$ for $Z = 0.001$, $\eta_{\rm AGB}=2$ for $Z=0.004$, and
$\eta_{\rm AGB}=4$ for $Z=0.008$, 0.02 and 0.04.  A lower value of
$\eta_{\rm AGB}$ implies a larger yield of carbon because lower mass
loss rates lead to longer stellar lifetimes and hence to more thermal
pulses.  As a consequence, more C is dredged up to the stellar surface
(see Chiappini et al. 2003b).
\item
For massive stars ($M >8 M_{\odot}$) we adopt the yields of Thielemann
et al. (1996, TNH) computed for the solar chemical composition.  With
this choice of stellar yields, the N from massive stars has a
secondary origin\footnote{A secondary element is a chemical species
  which is created from the metals originally present in the
  composition of a given star. In the Simple Model for chemical
  evolution (e.g. Matteucci 2001) it has been shown that the mass
  abundance in the gas of a secondary element evolves as the square of
  the total metallicity.}, whereas a fraction of the N coming from
low- and intermediate-mass stars has primary origin (vdHG).
\item
Finally, we use the yields of Nomoto et al. (1997, model W7) for SNIa.
These are assumed to originate from C-O white dwarfs in binary systems
that have accreted material from a companion (the secondary), reached
the Chandrasekar mass, and exploded via C-deflagration.
\end{enumerate}

In the next sections we will also modify the base model to include
other nucleosynthesis prescriptions.  For C, N, and O the above
fiducial model will be compared with models computed using the stellar
yields of the Geneva group, which include the effects of both mass
loss and rotation.  The adopted yields are the same as the ones
adopted by Chiappini et al. (2006a), where details can be found.
Briefly, the yields of Meynet \& Maeder (2002, hereafter MM02) were
adopted for metallicities larger than $Z=10^{-5}$ over the whole mass
range, assuming a rotational velocity of 300 km s$^{-1}$.  Below this
value we adopted the yields of Hirschi (2007, H07) for massive stars,
while keeping the MM02 calculations for low and intermediate mass
stars.  For SNIa we kept the same prescription as in PM04 model.

\begin{table}
\centering
\begin{minipage}{120mm}
\scriptsize
\begin{flushleft}
\caption[]{Summary of model properties: physical quantities}
\begin{tabular}{l|llllllll}
\hline
\hline
$M_{lum}$ 	&$R_{eff}$ &  $\nu$ 	    & $\tau$& $t_{gw}$     \\
({$M_{\odot}$}) & ({kpc})  &  ({Gyr$^{-1}$}) & (Gyr)& (Gyr)        \\
\hline
$10^{10}$       & 1        & 3              & 0.5  & 1.30      &               &                      \\ 
$10^{12}$       & 10       & 22             &0.2    &   0.44   &               &                      \\
\hline
\end{tabular}
\end{flushleft}
\end{minipage}
\end{table}

\begin{table}
\centering
\begin{minipage}{120mm}
\scriptsize
\begin{flushleft}
\caption[]{Summary of model properties: stellar yields}
\begin{tabular}{l|llllllll}
\hline
\hline
Model & yields for C,N                                   \\
      & low- and intermediate & massive stars \\
      &    mass stars         &                \\
\hline
PM04  &     vdHG                         & TNH            \\
\hline
Model I&    vdHG                         & H07 for $Z<10^{-8}$\\ 
       &                                 & MM02 otherwise      \\
\hline
Model IZ&   vdHG                         & H07 \\
\hline
Model II&   MM02                         & H07 for $Z<10^{-8}$\\ 
       &                                 & MM02 otherwise      \\                     
\hline
Model IIZ&  MM02                         & H07 for $Z<0.004$\\ 
       &                                 & MM02 otherwise      \\
\hline
\end{tabular}
\end{flushleft}
\end{minipage}
\end{table}

The different cases studied here, with different stellar evolution
prescriptions, are summarized below and in Table 2.

\begin{enumerate}
\item Model I: as PM04, but the { CNO} yields for massive stars ($m >
  8 M_{\odot}$) are as in Chiappini et al. (2006a).
\item Model IZ: as Model I, except that for CNO we adopt the yields
  for massive stars at $Z=10^{-8}$ (H07) as if they were valid for the
  entire metallicity range. The effect is to increase the output
  yields of C and N over those of Model I.  Although this is clearly
  not physically justified, we use this model to illustrate the effect
  of boosting C and N.
\item Model II: in this case, we adopted the nucleosynthesis
  prescriptions for CNO used in Chiappini et al. (2006a) for the full
  stellar mass range.  {\item Model IIZ: as Model II, but here the H07
    yields for massive stars are adopted for $Z<0.004$. This could
    apply to systems with high star formation rates such as
    ellipticals, where the number of fast rotators could have been
    larger, extending to larger metallicities than in the MW (see
    Decressin et al. 2007 for the same suggestion in the case of
    globular clusters)}.
\end{enumerate}

\section{Results}

In Table 3 we present the mass-weighted average stellar abundance
ratios (cf. Eq.~\ref{pp75}) predicted by each of the models described
in the previous section.  In the following sections, we analyse each
of the models in detail. We first focus on Mg and Ca, whose abundances
are well reproduced by the PM04 model without any further modification
of the stellar yields.  We then discuss C and N.

\begin{table}
\centering
\begin{minipage}{120mm}
\scriptsize
\begin{flushleft}
\caption[]{Mass-weighted stellar abundance ratios}
\begin{tabular}{l|llllllll}
\hline
\hline
\tiny
$M_{lum}$ 	& [$\langle$Mg/Fe$\rangle$] & [$\langle$Fe/H$\rangle$] & [$\langle$Ca/Fe$\rangle$] & [$\langle$C/Fe$\rangle$] & [$\langle$N/Fe$\rangle$]\\
({$M_{\odot}$}) &   \\
\hline
PM04\\
\hline
$10^{10}$       & 0.15    & 0.03   & -0.07  &  -0.55 &  -0.19\\   
$10^{12}$       & 0.30    & 0.16   & 0.04  &  -0.5   &  -0.18\\    
\hline
I \\
\hline
$10^{10}$       & 0.15    & 0.03   & -0.07  & -0.02 &  -0.05\\   
$10^{12}$       & 0.30    & 0.16   & 0.04  &  0.14   &  -0.01\\  
\hline
IZ\\
\hline
$10^{10}$       & 0.15    & 0.03   & -0.07  &  0.22 &   0.02\\   
$10^{12}$       & 0.30    & 0.16   & 0.04  &  0.34   &   0.09\\  
\hline
II\\
\hline
$10^{10}$       & 0.15   & 0.03   & -0.07  &  -0.002 &  -0.34\\   
$10^{12}$       & 0.30    & 0.16   & 0.04  &  0.154   & -0.19\\ 
\hline
IIZ\\
\hline
$10^{10}$       & 0.15   & 0.03   & -0.07  &  0.09 &  -0.27\\   
$10^{12}$       & 0.30    & 0.16   & 0.04  &  0.236  & -0.14\\ 
\hline
\end{tabular}
\end{flushleft}
\end{minipage}
\end{table}

\subsection{Mg and Ca}

\subsubsection{\rm [$\langle$Mg/Fe$\rangle$]}

The analysis of the mass-[$\langle$Mg/Fe$\rangle$] relation has been
comprehensively addressed in PM04, Pipino \& Matteucci (2006, 2008),
and Pipino et al. (2009), where a detailed comparison with the
observations has been made.  The Graves et al (2007) observational
result for this abundance ratio is remarkably similar to previous work
(e.g. Nelan et al. 2005, Smith et al. 2007, Bernardi et al. 2003),
therefore we do not repeat the analysis here. We just recall that an
overabundance of Mg relative to Fe is the key indicator that galaxy
formation occurred before a substantial number of Type Ia SNe could
explode and contribute to lower the [Mg/Fe] ratio (for the time-delay
model, see Matteucci 2001). In addition, the [$\langle$Mg/Fe$\rangle$]
ratio in the cores of ellipticals increases with galactic mass
(Worthey et al. 1992; Weiss et al. 1995; Nelan et al. 2005).  This
relation seems to be already in place at redshift 0.4 (Ziegler et al
2005).

In order to account for this trend in the star formation time-scale there are at least three possibilities that have been
discussed in the literature. 
One involves the loss of residual gas
via galactic winds that are initiated earlier in the most massive
objects (the \emph{inverse wind} picture, see Matteucci 1994).
Another possibility is to assume an initial mass function (IMF) which
becomes systematically flatter with increasing galactic mass.  A
selective loss of metals could also be the cause for the increase of
[$\langle$Mg/Fe$\rangle$] in more massive galaxies (see Matteucci et
al. 1998).  We consider the first possibility to be the best
motivated, and we will refer to it as \emph{chemical downsizing}.
Maiolino et al. (2008) have presented observational evidence
for chemical downsizing as far back as $z\sim 3$.  

Recently Pipino et al. (2009) showed that a more physically motivated
value for $\nu$ gives results in excellent agreement with the
observations (and with PM04). In particular, following Silk (2005),
they argue that a short ($10^6-10^7$ yr) super-Eddington phase can
provide the accelerated triggering of star formation needed.  The
subsequent quenching of star formation is accomplished by the SN
energy input and results in the usual black hole mass-spheroid
velocity dispersion relation (Magorrian et al. 1998).

The observational scatter in the relation between
[$\langle$Mg/Fe$\rangle$] and galactic mass can be entirely explained
as intrinsic scatter. Local effects, such as variations in the SN
feedback efficiency { of the order of a factor of two} with respect to
the best model case, can induce or delay the onset of the galactic
wind and thus contribute to setting the final value for
[$\langle$Mg/Fe$\rangle$].

On the other hand, such a relation cannot be explained by a sequence
of dry-mergers that create a massive spheroid starting from low-mass
$\alpha$-enhanced building blocks without violating other fundamental
constraints (Pipino \& Matteucci, 2008).

The PM04 best model predicts $d[\langle{\rm Mg/Fe}\rangle]/dlog\sigma
= 0.37$, in excellent agreement with the reported value of 0.36 by
Graves et al. (2007). This abundance ratio was used by PM04 as a
crucial constraint {in order to set all the relevant physical
  quantities reported in Table 1. The same values are adopted here so
  that Models I, IZ, II and IIZ also obey to this fundamental
  constraint.}

In the following sections we study other abundance ratios and, when a
discrepancy is found between model predictions and the observations, 
{ the problem is addressed by} comparing {model predictions computed with different} stellar yield prescriptions.

\subsubsection{\rm [$\langle$Ca/Fe$\rangle$]}

The observations show that, although Ca belongs to the group of the
$\alpha$ elements, the strength of the observed Ca lines appear to
follow [$\langle$Fe/H$\rangle$] instead of [$\langle$Mg/H$\rangle$]
(Worthey, 1998; Trager et al., 1998; Saglia et al. 2003).  In
particular Thomas et al. (2003) suggested [$\langle$Ca/Mg$\rangle$]
=$-$0.15, and Saglia et al. (2003), after a detailed analysis of
several possible sources of error, claimed that the Ca depletion in
ellipticals is real.  Such a result is confirmed by Graves et
al. (2007), who find the [Ca/Fe] ratio to be nearly solar over the
entire galactic mass range.

The PM04 calculations already suggested [$\langle$Ca/Mg$\rangle$]
$=-$0.152 and [$\langle$Ca/Fe$\rangle$] $=-0.03$ (these values are for
the core of a $10^{11}M_{\odot}$ PM04 galaxy) and are in excellent
agreement with the above mentioned observations.  Here we add that at
either lower or higher galactic masses the predicted ratio is close to
the solar value with a very mild relation with mass.  A linear regression
fit of the mass weighted stellar abundance ratio as a function of the
stellar velocity dispersion returns $d[\langle{\rm
    Ca/Fe}\rangle]/d\log\sigma = 0.27$ for PM04 (and hence the same
value for Models I, IZ, II and IIZ), steeper than the slope of 0.13
reported by Graves et al. (2007).  We notice that if we adopt Woosley
\& Weaver (1995) yields for massive stars we obtain $d[\langle{\rm
    Ca/Fe}\rangle]/d\log\sigma = 0.18$, which is in better agreement
with the observed slope but produces [$\langle$Ca/Fe$\rangle$] values
offset about 0.15 dex higher than observations at a given mass.

As shown in PM04, this result can be explained simply by SN yields. { In fact,}
the $\alpha$-elements exhibit different degrees of enhancement
with respect to Fe { in the results of the chemical evolution
model (see Fran\c cois et al. 2004)}. This is due, from a theoretical point of view, to the different
degree of production of each element in Type II and Ia SNe.  In
particular, Ca and Si show a lower overabundance relative to Fe than O
and Mg, since they are also produced substantially in Type Ia SNe. 
In fact, the mass of Ca ejected during a SNIa explosion in
the model W7 (Nomoto et al., 1997) is $\sim 0.012 M_{\odot}$, whereas
the contribution of a typical Type II SNe (averaged over a Salpeter
IMF in the mass range $10 - 50 M_{\odot}$) is $\sim 0.0058 M_{\odot}$
(see Table 3 of Iwamoto et al., 1999). While some fine tuning of the
stellar nucleosynthesis could improve even further the agreement with
observations, we stress that PM04 explains the behaviour of the Ca
abundance as a function of mass without requiring changes in the
adopted chemical evolution model parameters.

\subsection{C and N}

\begin{figure}
\includegraphics[width=9cm,height=9cm]{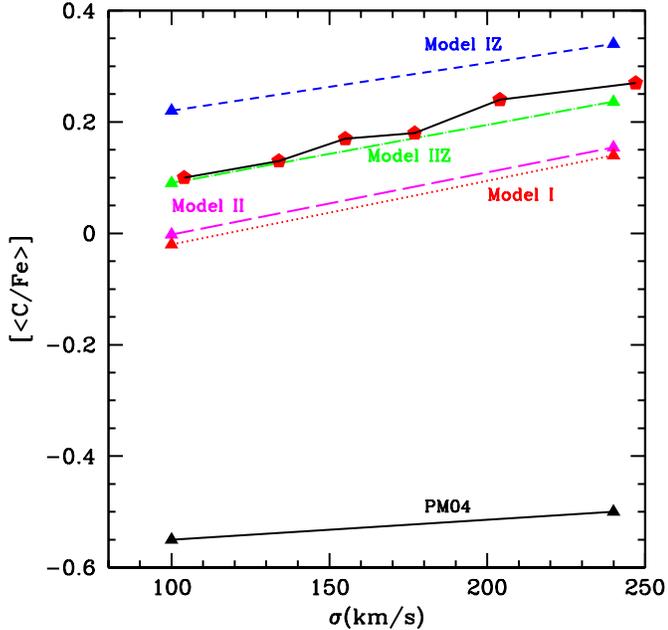}
\caption{Mass-weighted [$\langle$C/Fe$\rangle$] as a function of
  galactic velocity dispersion predicted by PM04 (triangles linked by
  a solid line) and Models I (dotted line), IZ (short-dashed line), II
  (long-dashed line) and IIZ(short dash-dotted line) compared to the
  observed relation found by Graves et al. (2007, pentagons linked by
  a solid line).
\label{fig1}}
\end{figure}

Figs.~\ref{fig1} and ~\ref{fig2} show our model predictions for the
[$\langle$C,N/Fe$\rangle$]-mass relations compared with the observed
ones by Graves et al. (2007) {(represented by pentagons connected by a solid line}.  The PM04 Model (triangles joined by a
solid line) predicts both [$\langle$C/Fe$\rangle$] and
[$\langle$N/Fe$\rangle$] to be below solar and to show no dependence
on galactic mass, at variance with the observations. The other models (i.e. I:
dotted line; IZ: dashed line; II: long-dashed line; IIZ: long-dashed
dot line), while keeping the same star formation history as PM04, show
different trends in abundance ratios with mass, reflecting the
different assumptions about the nucleosynthesis. Most of the modified
models lead to an improvement over the predictions of PM04.  Thus, for
the first time we can use local elliptical galaxies as a test bench
for stellar nucleosynthesis studies.  In the following sections will
we discuss the C and N results in further detail.

\subsubsection{\rm [$\langle$C/Fe$\rangle$] { in the integrated stellar spectrum}}

We first focus on the predictions of our new models concerning the
[$\langle$C/Fe$\rangle$] abundance ratio in Fig~\ref{fig1}.  Model I
(dotted line) predicts $d[\langle{\rm C/Fe}\rangle]/d\log\sigma =
0.4$, in good agreement with the value of 0.5 reported by Graves et
al. (2007, solid line). However, this model has a zero-point of the
mass-[$\langle$C/Fe$\rangle$] relation $\sim$0.1 dex smaller than the
observed one.  

The large difference between Model I and PM04 is due to
the larger contribution of massive stars to the carbon production in
the case of the Geneva group yields with respect to the TNH ones. The
larger C yields of the Geneva group are found not only at very low
metallicities (due to rotation, as explained before) but also at solar
metallicities (see Fig.1 of Chiappini et al. 2003b for a comparison of
the C yields of the two groups at $Z=Z_{\odot}$). In fact, the
mass-loss rate, which increases with metallicity, 
contribute to increase the overall C yields (see MM02).

A similar result is found for Model II (long-dashed line), which
differs from Model I only in the C yields for intermediate mass stars
(vdHG in Model I and MM02 in Model II). The difference between these
two models is expected to be small since, as shown by Chiappini et
al. (2003a) the yields of C from vdHG and MM02 for low- and
intermediate-mass stars are similar. In fact, although MM02 yields do
not include the third dredge up contribution, they obtain a similar
effect thanks to rotation and mass-loss (see Chiappini et al. 2003a
for details).

Model IZ (dashed line) predicts: $d[\langle{\rm
    C/Fe}\rangle]/d\log\sigma = 0.3$ with a larger zero-point. In this
case the model galaxies are more C-enhanced than the observed ones.
{ This is not surprising, as Model IZ assumes fast rotators to exist at all metallicities
and not only below Z=10$^{-8}$ as done in Chiappini et al. (2006a).}
This illustrates the effect of substantial
additional C enhancement from massive stars but is not
based on any physical motivation.  

Finally, Model IIZ matches the [$\langle$C/Fe$\rangle$]-mass relation
quite well. In this case, the C contribution from fast rotators is
extended up to $Z = 0.004$ and is therefore larger than in Model II
(where fast rotators exist only up to $Z = 10^{-8}$, see Section 3.3).

Overall, it is encouraging that all of the models which include the
effects of rotation { and mass loss} in massive stars (Models I, IZ, II, and IIZ)
produce reasonable agreement with the data, showing the correct slope
of the mass-[$\langle$C/Fe$\rangle$] relation, with zero-point offsets
of the order of only $\pm0.1$ dex from the data.  There are variations
depending on the various yield tables adopted, but these variations
are relatively small.  This suggests that modelling of the
nucleosynthesis of C in massive rotating stars might approach a
consensus soon and therefore that C abundances will be a powerful new tool
for constraining galaxy star formation histories.

It should be kept in mind that the transformation of Lick indices into
mean stellar abundances carries uncertainties, as discussed in Section
2.2.  These uncertainties are more likely to affect the zero-point
calibration of the abundances, while the {\it relative} abundances and
therefore the abundance trends with $\sigma$ should be fairly robust.
Finally we notice that both our model predictions and the data set of
Graves et al. (2007) are at odds with the recent claim (Clemens et
al. 2006) of a decrease in the [$\langle$C/H$\rangle$] value in stars
as function of the galactic mass.

\subsubsection{\rm [$\langle$N/Fe$\rangle$] in the integrated stellar spectrum}

\begin{figure}
\includegraphics[width=9cm,height=9cm]{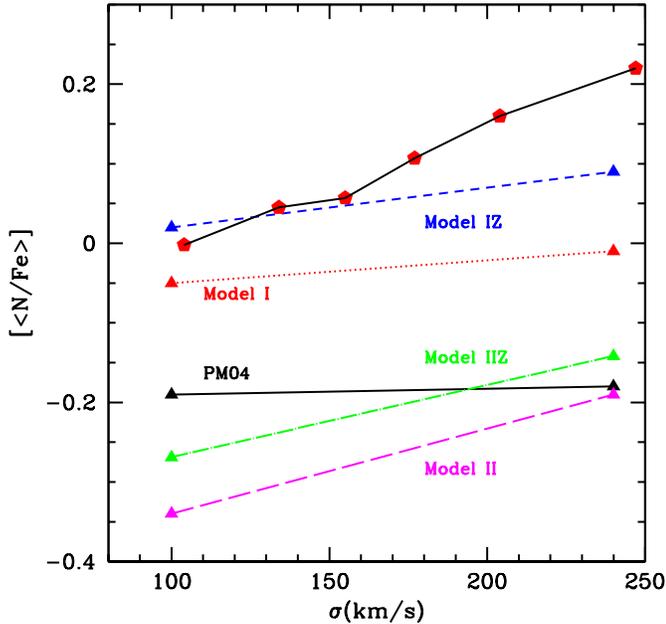}
\caption{Mass-weighted [$\langle$N/Fe$\rangle$] as a function of
  galactic velocity dispersion predicted by PM04 (triangles linked by
  a solid line) and Models I (dotted line), IZ (dashed line), II
  (long-dashed line) and IIZ (short-dashed dot line) compared to the
  observed relation found by Graves et al. (2007, pentagons linked by
  a solid line).
\label{fig2}}
\end{figure}

Concerning N, models I and IZ (see Fig~\ref{fig2}, symbols as in
Fig.~\ref{fig1}), substantially increase the zero-point of the
mass-[$\langle$N/Fe$\rangle$] relation with respect to PM04, hence
improving the agreement with the observations. However, both models
predict a rather flat slope: $d[\langle{\rm N/Fe}\rangle]/d\log\sigma
= 0.1$ (a factor of 5 flatter than the value indicated by the SDSS
galaxies). We remind the reader that Models I and IZ assume primary N
production in massive stars, whereas in PM04 the N from massive stars
is only secondary.

On the other hand, Model II is able to predict a slope ($d[\langle{\rm
    N/Fe}\rangle]/d\log\sigma = 0.4$) for the
mass-[$\langle$N/Fe$\rangle$] relation, in much better agreement with
Graves et al.'s findings, but it does not improve upon the original
PM04 model in terms of the absolute value. In fact,
[$\langle$N/Fe$\rangle$] is under-predicted by 0.4 dex (i.e. more than
a factor of 2) in this case.  The only differences between models I
and II are the adopted yields for low- and intermediate-mass
stars. Model I was computed with the yields of vdHG, which predict
larger quantities of N than the yields of MM02 (adopted in Model
II). These large N quantities are responsible for both the larger zero
point and the flatter behavior of the mass-[$\langle$N/Fe$\rangle$]
relation.  When assuming that the fast rotators play a role up to
metallicities $Z=0.004$ (Model IIZ), the [$\langle$N/Fe$\rangle$] zero
point is increased by $\sim$0.1 dex, which is still not enough to
account for the observed values.

In the case of Models II and IIZ, the differences between observed and
predicted values at a given galactic mass are probably too large to be
explained in terms of either uncertainties in the calibration of the
Lick system (see Section 2.2), or in the actual measurements. This
offset is larger than the 3$\sigma$ measurement error quoted in Graves
et al. (2007) for the observed zero-point of the
[$\langle$N/Fe$\rangle$]-$\sigma$ relation.

{ It is worth noting that, unlike in the case with
[$\langle$C/Fe$\rangle$], there is substantial disagreement in the
predicted values of [$\langle$N/Fe$\rangle$] between the four models,
both in the predicted values of $d[\langle{\rm
    N/Fe}\rangle]/d\log\sigma$ and in the zero-points (at the level of
0.4 dex).  This indicates that the various nucleosynthesis yield
predictions for N vary significantly between groups and are highly
sensitive to the mass and metallicity range over which rotation is 
assumed to be relevant.  The nucleosynthetic predictions
for N are therefore significantly less secure than those for C.}

Also, as discussed in Section 2.2, stellar [$\langle$N/Fe$\rangle$]
abundance ratio measurements in the observed galaxies are more uncertain
than the other abundance determinations.  

\subsubsection{\rm N abundance in the gas: model ellipticals}

\begin{figure*}
\includegraphics[width=8cm,height=8cm]{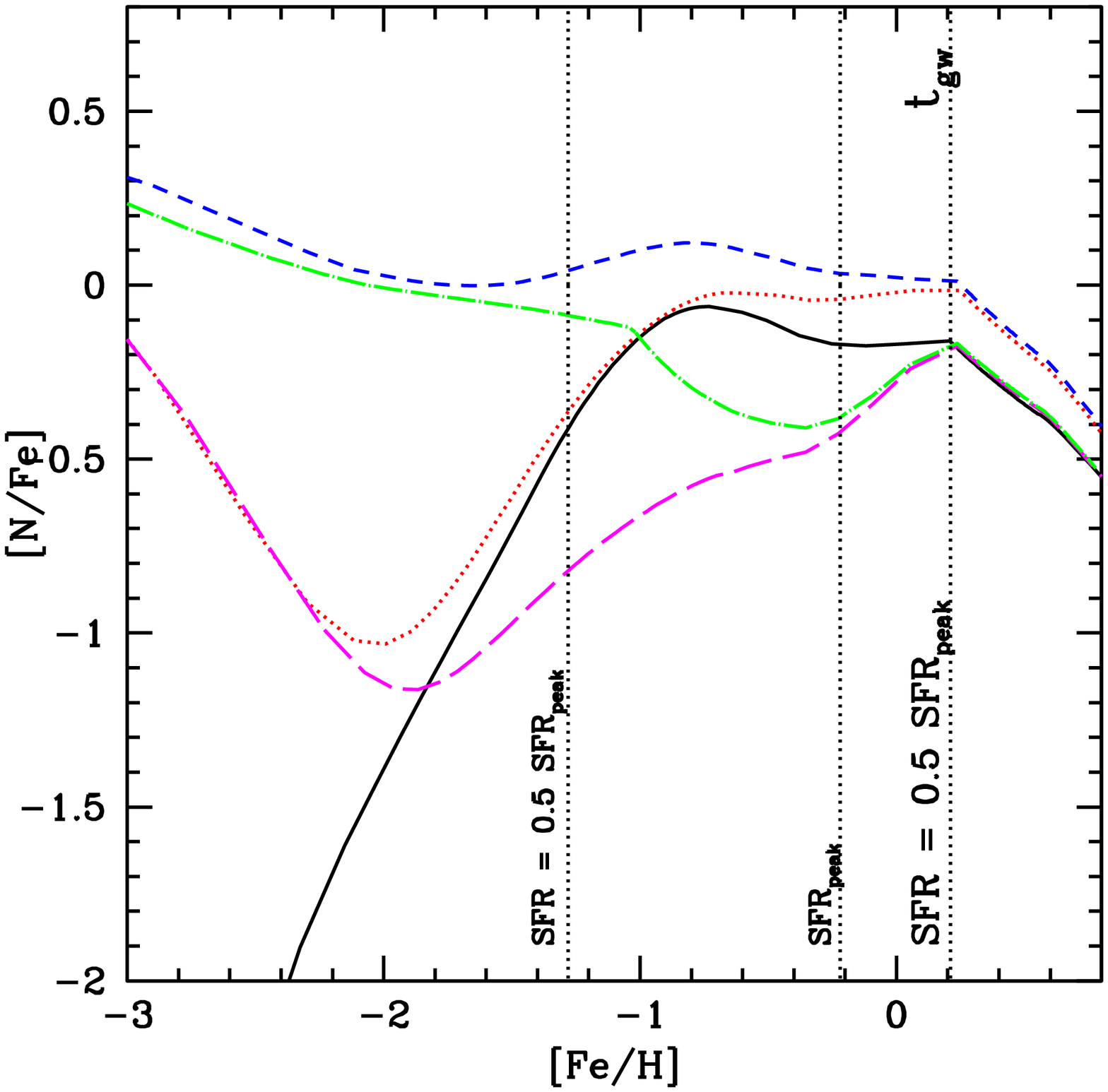}
\includegraphics[width=8cm,height=8cm]{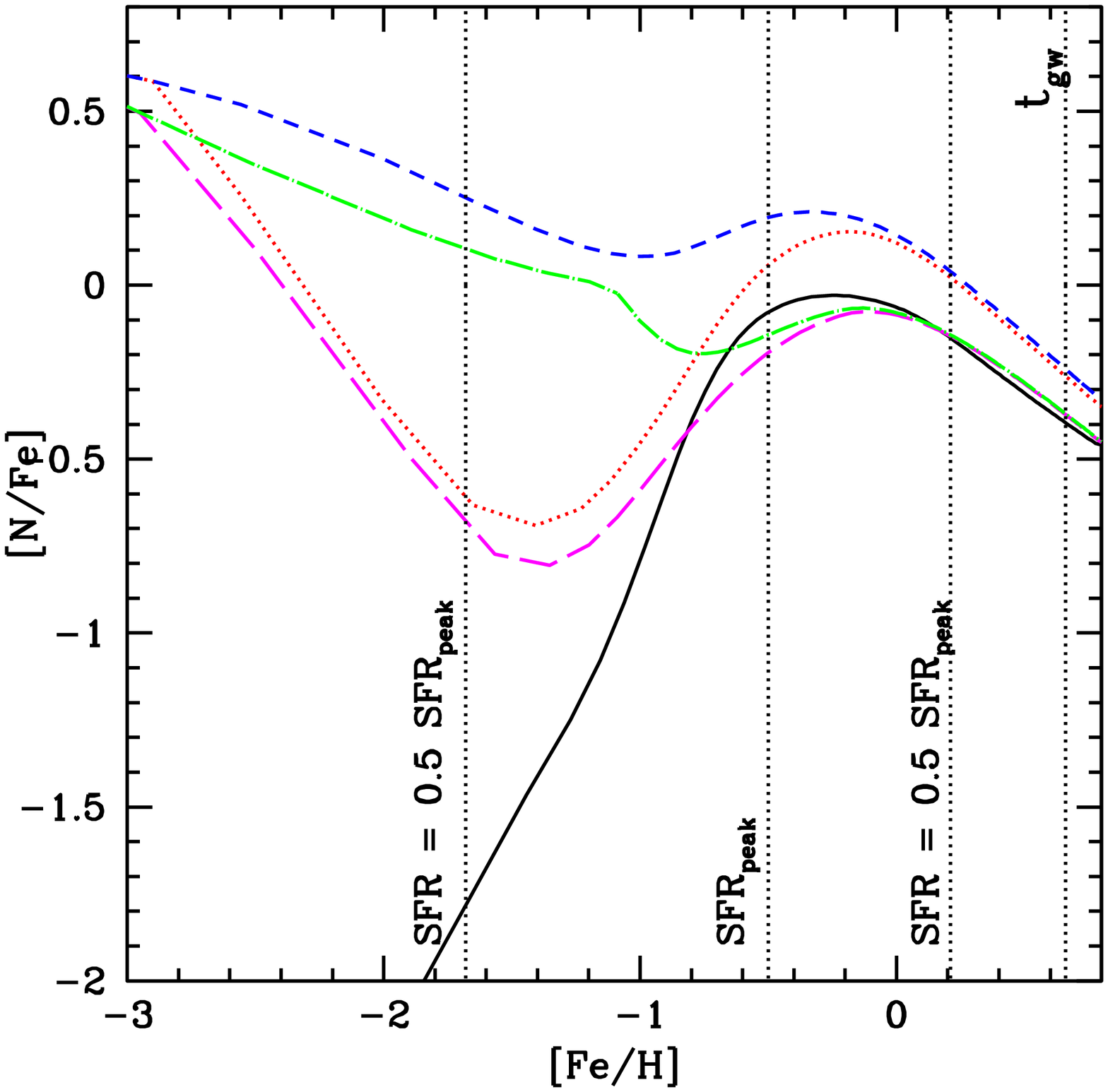}
\caption{[N/Fe] as a function of [Fe/H] \emph{in the gas} for a
  10$^{10}$ M$_{\odot}$ (left panel) and a 10$^{12}$ M$_{\odot}$ (right panel)
  elliptical galaxy.  Different lines represent the different models
  in Table 2: PM04 (solid line) and Models I (dotted line), IZ (dashed
  line), II (long-dashed line) and IIZ (short-dashed dot line).
  Vertical dotted lines indicate the metallicity where the SFR reaches
  its peak, as well as when it rises to and drops by a factor of two
  from this value.  Also shown is the metallicity at the onset of the
  galactic wind (labelled as $t_{gw}$), indicating the time (hence the
  metallicity) at which the star formation stops.  This illustrates
  the values of [N/Fe] at which most of the stars in each galaxy are
  formed.
\label{fig4}}
\end{figure*}

Since the predicted values for [$\langle$C,N/Fe$\rangle$] are the
result of the convolution of the temporal evolution of [C,N/Fe] in the
gas with the star formation rate, it is important to study the gas
abundances in order to understand the differences between the models
presented in the previous section.  In Fig.~\ref{fig4} we plot the
evolution of [N/Fe] in the gas phase of the $10^{10} M_{\odot}$ and
the $10^{12} M_{\odot}$ model galaxies (left and right panels,
respectively) as a function of [Fe/H].  The dotted lines mark relevant
times during the course of star formation in the galaxy, including the
metallicity at the time when the SFR reaches its peak, the
metallicities at which the SFR is within a factor of two from the peak
value, and the onset of the galactic wind (labelled as $t_{gw}$). This event
sets the time (hence the metallicity) at which the star
formation stops.  This illustrates the values of [N/Fe] at which most
of the stars in each galaxy are formed. 

First of all, a comparison of Figs~\ref{fig2} and~\ref{fig4} shows
that, even if N is more enhanced in more massive (hence more metal
rich) galaxies, this does not imply that N has a secondary origin!  It
is true that models in which N is mostly secondary (e.g., PM04) have
gas phase [N/Fe] abundances which increase steeply with [Fe/H], as
shown in Figure~\ref{fig4} (PM04 is the solid black line).  However,
this does not necessarily translate into a strong
[$\langle$N/Fe$\rangle$]-mass relation (see Figure~\ref{fig2}).  In
contrast, models which produce a better match to the observed
[$\langle$N/Fe$\rangle$]-mass relation (e.g., Model IZ, dashed blue
line), include substantial primary N production in massive stars and
show high gas-phase [N/Fe] abundances at low [Fe/H]
(Figure~\ref{fig4}).  

By comparing the left panel of Figure~\ref{fig4} ($10^{10} M_{\odot}$
case) to the right one ($10^{12} M_{\odot}$ case), it is clear that
the more massive galaxy features a higher [N/Fe] ratio in the gas
during the phases when the star formation proceeds at its maximum
rate.  Stars form out of this enriched material until the onset of the
galactic wind at time $t_{gw}$.  The difference between the $10^{10}
M_{\odot}$ galaxy and the $10^{12} M_{\odot}$ galaxy is the smallest
for the PM04 model, hence we predict a rather flat slope in the
[$\langle$N/Fe$\rangle$]-mass relation.  The difference is largest in
models II and IIZ, which therefore exhibit the steepest slope in the
[$\langle$N/Fe$\rangle$]-mass relation { (Fig. 2)}.

\begin{figure*}
\includegraphics[width=\textwidth,height=8cm]{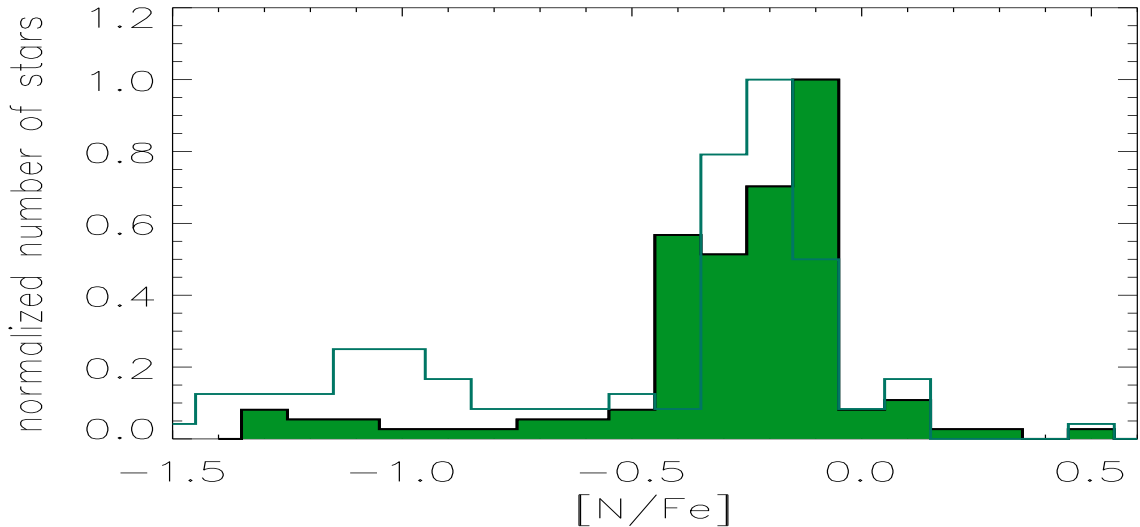}
\caption{Normalized number of stars as a function of [N/Fe] for model
  IIZ in the case of the $10^{10}M_{\odot}$ model galaxy (empty
  histogram) and of the $10^{12}M_{\odot}$ model (filled histogram).
\label{fig3bis}}
\end{figure*}

This result is quantified in Fig.~\ref{fig3bis} for Model IIZ: the
normalized distribution of stars as a function of [N/Fe] peaks at
higher values of the abundance ratio and features a smaller tail at
[N/Fe] $< -1$ in the $10^{12}M_{\odot}$ model galaxy (full histogram)
than in the $10^{10}M_{\odot}$ galaxy (empty histogram).  Note that
the values of [$\langle$N/Fe$\rangle$] derived by means of Eq.(3) and
shown in Table 3 correspond to the \emph{mean value} of distributions
like the one shown in Fig.~\ref{fig3bis}. The mean does not always
coincide with the peak value (i.e. the \emph{mode}) of the
distribution. Given the asymmetry of the { distributions}, the mean, in fact,
tends to be $\sim$0.1-0.2 dex lower than the mode. By comparing the
distribution of stars as a function of [N/Fe] for
different models, we can also explain the differences in the predicted
[$\langle$N/Fe$\rangle$] \emph{at a given mass} (Section 4.2.2). For
instance, model IZ features a distribution which peaks at higher
[$\langle$N/Fe$\rangle$] and has a larger mean value than model
PM04.  Finally, we note that all the models shown 
in Fig. 3 (with the exception of models IZ and IIZ) 
exhibit a [N/Fe] ratio below solar during most of the galaxy evolution.

Intermediate mass stars (i.e., $M < 8 M_{\odot}$) do not contribute
metals to the ISM until $\sim30$ Myr after the beginning of star
formation.  This corresponds to $Z=5\cdot10^{-4}$ in the $10^{10}
M_{\odot}$ galaxy and $Z=3\cdot10^{-3}$ in the $10^{12} M_{\odot}$
galaxy, where the evolution is faster.  Due to the very short
time-scale of galaxy formation (see Table 1), the lowest stellar mass
which can contribute to the chemical enrichment of the model galaxies
is $\sim 1.6 M_{\odot}$ in the $10^{10} M_{\odot}$ case and $\sim 2
M_{\odot}$ in the $10^{12} M_{\odot}$ case.  The differences in the N
production at low metallicities (i.e. [Fe/H] and [O/H] $<-$2) will { have little effect in the final}
 [$\langle$N/Fe$\rangle$] ratio because very few stars are
formed at these metallicities.

\subsubsection{\rm N abundance in the gas: high redshift progenitors}

Ideally, one would use gas phase abundance ratios as shown in
Fig~\ref{fig4} to further refine the model. Unfortunately, no such
measurements exists in low-redshift ellipticals, as they are almost 
devoid of gas. In order to perform such an exercise we must resort to
high-redshift objects, such as the Lyman Break Galaxy (LBG)
MS1512-cB58 (Pettini et al. 2001, Teplitz et al. 2000).  This is
currently the brightest LBG known, because it is gravitationally
lensed.  The LBG is at $z=2.7672$ and has a luminous mass of $\sim
10^{10}M_{\odot}$ with a star formation rate of $\sim 40 \,\rm
M_{\odot}yr^{-1}$ (Pettini et al. 2001). We use this galaxy because,
from the analysis of its properties and its chemical abundance
pattern, Matteucci \& Pipino (2002) suggest MS 1512-cB58 to be a
low-mass young elliptical experiencing its main episode of star
formation and galactic wind.

\begin{figure}
\includegraphics[width=9cm,height=9cm]{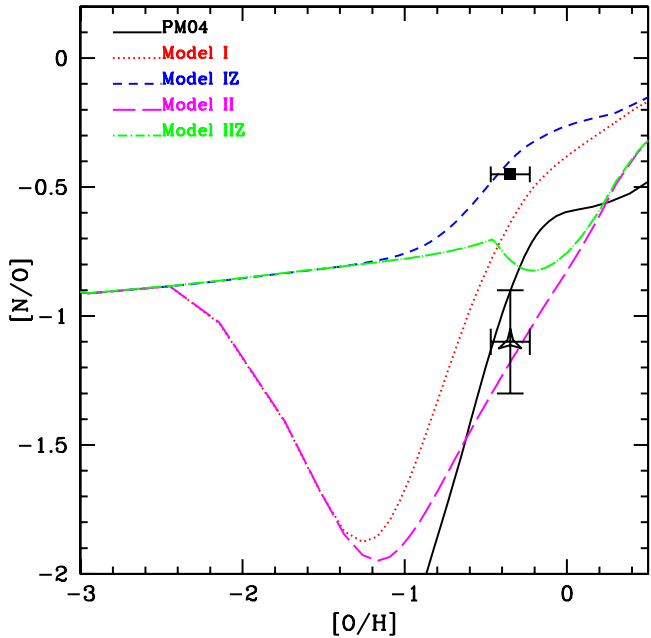}
\caption{[N/O] as a function of [O/H] \emph{in the gas} for the
  different $10^{10} M_{\odot}$ models: PM04 (solid line) and Models I
  (dotted line), IZ (dashed line), II (long-dashed line) and IIZ
  (short-dashed dot line).  The triangle with error-bars is the observed
  value for the Lyman Break Galaxy MS1512-cB58 (Pettini et al 2001), whereas the
solid square is the estimate from Teplitz et al. (2000).}
\label{fig3}
\end{figure}

Figure~\ref{fig3} shows the evolution of the [N/O] abundance ratio as
function of [O/H] \emph{in the gas} for our model galaxy with
$M_{\star} = 10^{10}$M$_{\odot}$, for each of the nucleosynthesis
cases studied in the previous sections. In this figure, the point with
error-bars represents the LBG\footnote{Note that the N abundance is
  inferred from absorption lines (Pettini et al., 2001), whereas the O
  abundance comes from emission line measurements (Teplitz et al.,
  2000)}.  The models shown in Fig.~\ref{fig3} have star formation
rates similar to the one observed in MS1512 cB-58. The solid line
represents the original PM04 model.  The [N/O] abundance ratio rises
quite steeply with metallicity\footnote{O contributes 60\% of the
  total metallicity in this system.} because, in the PM04 model, the N
from massive stars is assumed to be a secondary element. The
predictions of the PM04 model (which is an improved version of the
chemical evolution model presented by Matteucci \& Pipino, 2002) are
within 1$\sigma$ of the observed value.

Model II (long dashed line) performs equally well, exhibiting an early
(i.e. [O/H] $< -$3) plateau with values around [N/O] $= -1$, followed
by a decrease due to the metallicity-dependent yields of massive
stars.  Later, intermediate-mass stars start contributing to the
chemical enrichment (once [O/H] $> -$1 dex in the case shown in
Fig.~\ref{fig3}). Such a contribution starts to be important around
[O/H] $= -0.5$, where a new rise in the [N/O] ratio is observed.

The detailed behaviour of the curve predicted by Model II is still
uncertain, primarily due to the following reasons: (a) There are still
no complete grids of stellar yields for models with rotation. Here we
are combining the calculations of MM02 for 300 km s$^{-1}$ with the
more recent calculations of H07, which includes the fast rotators
(500--800 km s$^{-1}$).  Grids with intermediate values of rotational
velocities at different metallicities are currently not available.
Moreover, there are small differences between MM02 and { H07}
calculations, such as different amounts of convective overshooting,
that prevent a completely self-consistent approach. (b) The
contribution to N { by} Super-AGB stars\footnote{Super-AGB stars are
  massive enough to ignite C in a degenerate core but do not proceed
  into neon burning and beyond.  They are thus an intermediate case
  between intermediate-mass and massive stars.} could be important
(Siess 2007).  This is not included here as no stellar yield
calculations for these objects are available at the moment. These
stars would play a role at earlier times than low- and
intermediate-mass stars and could contribute to fill up the valley
seen in Fig.~\ref{fig3} (also contributing to increase the median
value of [$\langle$N/Fe$\rangle$] in Fig.~\ref{fig2}). (c) In MM02,
the intermediate-mass stars were not followed up to the end of their
evolution and the contribution of hot bottom burning could be missing
(but see Chiappini et al. 2003b).

Despite of the above caveats, the important point here is that Model
II predicts a high N/O ratio at early phases that later decreases and
then increases again. The amount of the decrease in N/O at
intermediate metallicities is still uncertain but it clearly shows
that some metallicity-dependence of the N yields in massive stars is
necessary in order to bring [N/O] back up to levels consistent with
the Lyman Break galaxy.

On the other hand, Models IZ and IIZ, which are built to enhance C and
N in order to fit the \emph{stellar} ratios in local ellipticals,
predict a large N abundance with respect to O at low [O/H].  This is
because we let massive stars produce N with the same stellar yields as if they were in
the lowest metallicity regime.  { The sudden downturn of Model IIZ at
[O/H] $\sim -0.5$ is due to the adopted threshold metallicity for
including the fast rotators.  This drop is much less important than
what happens in Model II, where the fast rotators contribute during a
much shorter period of the evolution of these galaxies.  In other
words, in Model IIZ the drop is just shifted to larger metallicities
because we allowed fast rotators to contribute for longer times (up to
larger metallicities). } 

Model I (dotted line) is an intermediate case between PM04 and Model
II.  At variance with the fiducial PM04 model, it features an early
enhancement of N owing to the production from low-metallicity massive
stars, as in Model II.  It also shows a steeper rise around [O/H]
$=-0.5$ due to the fact that in this case, vdHG nucleosynthesis for
intermediate mass stars is producing more N (the { hot bottom burning}
contribution).

While Model I and IIZ are still marginally consistent with the data
from Pettini et al. (2001, triangle in Fig.~\ref{fig3}), Model IZ---which better matches the
[$\langle$N/Fe$\rangle$]-mass relation---predicts a [N/O] ratio more
than a factor of three larger than the observed LBG value at [O/H]
$=-0.35$.  However, the gas-phase N abundance inferred from
\emph{emission} lines in the same galaxy is [N/O] $\sim -0.5$ (Teplitz
et al. 2000, full square in Fig.~\ref{fig3}), more than three times higher than the [N/O] inferred
from the absorption line measurements of Pettini et~al. (2001).
Pettini et~al. argue that the N abundance inferred from emission lines
may be substantially overestimated.  Nonetheless, we can interpret the
factor of three difference in the N abundance between \emph{emission}
and \emph{absorption} lines as an estimate of the uncertainty in the
gas-phase measurements.  Using the Teplitz et~al. (2000) emission line
abundance brings Model IZ into excellent agreement with \emph{both}
stellar- and gas-phase abundances (full square in Fig.~\ref{fig3}). Unfortunately,
they do not provide errorbars, therefore we cannot quantitavely estimate
the level of agreement with our model prediction.

It is important to stress that a comparison with a single
high-redshift object, which my be a young elliptical caught in the act of 
formation, may not be appropriate.
More data on gas { phase} abundances in high redshift objects are needed to
confirm the tension emerging with respect to abundances measured in
stars and to reconcile \emph{emission}- and \emph{absorption}- line
measurements of the gas-phase abundance.  However, along with the
observations of DLA (which also show systematically lower N/O ratios
than the stars in the MW, again suggesting a $Z$-dependency of the N
yields in massive stars), such a comparison demonstrates the existence
of a dichotomy between stellar abundances (which tend to favor higher
N abundances) and gas phase abundances.

{ Our analysis demonstrates a potential inconsistency between the
stellar N abundances observed in local ellipticals and the gas-phase
abundances observed in a likely progenitor of these galaxies at higher
redshift.  The stellar [$\langle$N/Fe$\rangle$] abundance ratio
appears to require substantial N enrichment due to both massive and
intermediate-mass stars (e.g., Model IZ), but this model produces an
[N/O] abundance ratio that is too high to match the LBG data from
Pettini et al. (2001).  This discrepancy disappears if the [N/O]
abundance ratio of MS1512-cB58 is in fact [N/O] $= -0.5$, as
determined by Teplitz et~al. (2000).

Another possible solution to this discrepancy is that the
[$\langle$N/Fe$\rangle$] ratios inferred from stellar spectra are
overestimated by about 0.3 dex (this is the amount necessary to bring
it into agreement with model prediction and the Pettini et~al. LBG
data at the same time).  The various uncertainties in the
[$\langle$N/Fe$\rangle$] measurements were discussed in Section 2.2.
An offset of 0.3 dex is larger than can easily be accounted for in
this way, but is certainly not impossible, given the uncertainties in
the modelling process, the paucity of good test data, and the fact
that these constitute the first attempts to quantify
[$\langle$N/Fe$\rangle$] in unresolved stellar populations.}

\section{Conclusions}

In this paper we have analysed the mass-[Ca,C,N/Fe] ratio relations
observed in local spheroids by means of a chemical evolution model
which successfully reproduces the mass-[$\langle$Mg/Fe$\rangle$] and
the mass-metallicity relations (the PM04 model). Our theoretical
predictions are compared to the observed relations inferred from
integrated spectra of early-type galaxies in SDSS by Graves et
al. (2007). We find that:

\begin{itemize}
\item The [$\langle$Ca/Fe$\rangle$]-mass relation is naturally
  explained by the PM04 best model without any further assumption
  needed.  In particular, the under-abundance of Ca with respect to Mg
  can be attributed to the relative contributions of SNe Type Ia and
  Type II to the nucleosynthesis of these two elements.

\item The standard nucleosynthesis prescriptions adopted by PM04
  (i.e. where rotation and mass loss are not taken into account) lead
  to chemical evolution models in which C traces Fe and N behaves as a
  secondary element (i.e., there is no primary nitrogen contribution
  from massive stars).  Instead, the abundance ratios measured for the
  stellar population in local spheroids suggest that both C and N behave
  more like $\alpha$-elements (e.g., Mg) in the sense that the
  [$\langle$C,N/Fe$\rangle$] abundance ratios are super-solar and
  increase with the galactic mass.

\item We show that in order to reproduce the [$\langle$C/Fe$\rangle$]-
  and [$\langle$N/Fe$\rangle$]-mass observed relations it is necessary
  to consider new nucleosynthesis prescriptions, which take into
  account both the effects of rotation and mass-loss (Meynet et
  al. 2009). These two important physical processes have already
  proven to be crucial for our understanding of the abundance ratios
  in metal-poor stars in the Galactic halo (Spite et al. 2005,
  2006). Indeed, the high N/O, C/O and low $^{12}$C/$^{13}$C ratios
  observed in stars with [Fe/H] $< -$2.5 suggest the existence of fast
  rotating stars in the very early Universe (Chiappini et al. 2006a,b,
  2008).

\item The various different prescriptions for the effects of rotation
  and mass loss on the nucleosynthesis of C considered here all
  produce results that are consistent with the data and
  with each other.  This indicates that the nucleosynthesis of C is
  now understood well enough { and hence} C abundances can be used to place
  constraints on chemical evolution in galaxies.  The same level of
  consensus is not found for N abundances from the various models.
  This suggests that the nucleosynthesis of N is more sensitive to the
  differences in stellar evolution models and that the existing yield
  tables for N are perhaps less reliable than those for C.

\item Once these new stellar yields are implemented in the PM04
  chemical evolution model, a remarkable agreement with the
  observations is attained (within the observational uncertainties)
  both for the slope and the zero-point of the
  mass-[$\langle$C/Fe$\rangle$] relation (Model II). This shows that
  significant amounts of C must be produced in massive stars (and not
  only in intermediate mass stars) in order to reproduce the
  [$\langle$C/Fe$\rangle$] abundances of massive early type galaxies.

\item The same model reproduces the steep slope observed for the
  mass-[$\langle$N/Fe$\rangle$] relation in the local universe. This
  model also provides predictions for N/O ratios in the gas phase
  abundance of young spheroids.  The models considered here cannot
  simultaneously predict the high values of [$\langle$N/Fe$\rangle$]
  observed in local stellar populations and the low values of [N/O]
  observed in the gas-phase of LBG MS1512-cB58.  A factor of $\sim0.3$
  dex is required to resolve this difference.  There are significant
  uncertainties in both the stellar and gas-phase measurements of N,
  as well as the { N} yields, all of which may contribute
  to the observed discrepancy.

\end{itemize}

In summary the PM04 model, updated to include new nucleosynthetic
yields from stellar models which account for stellar rotation and mass
loss, can account for the [$\langle$Mg,Ca,C/Fe$\rangle$]-mass relations
observed in local spheroids.  Discrepancies may remain in the case of
N.  However, our best model (Model II) might be completely reconciled
with observations if \emph{emission}-line gas phase N abundance of MS
1512-cB58 is considered, rather than the N abundance determined from
\emph{absorption}-line measurements.  Another way to reconcile the
discrepancy is if the [$\langle$N/Fe$\rangle$] in the galaxies,
inferred from Lick indices, is overestimated by $\sim 0.3$dex.  If this
turns out to be the case, our results suggest that fast stellar
rotation at low $Z$ and large mass-loss rates at higher metallicities
in massive stars have an important impact on the chemical evolution of
early-type galaxies, as has already been shown for the Milky Way. On
the other hand, if the [$\langle$N/Fe$\rangle$] ratios inferred for
the SDSS galaxies are correct and if the \emph{absorption}-line gas
phase N abundance of MS 1512-cB58 are typical of young spheroids, a
tension between observed gas and stellar abundance ratios remains,
suggesting that further processes not envisaged in the present work
should be taken into account.  However, the uncertainties associated with
both stellar and gas-phase N abundance measurements, as well as the
uncertainties in the nucleosynthesis of N do not allow us to draw firm
conclusions. 

Finally, we note that, thanks to large sample sizes, high quality data
and improved stellar population models, we are able for the first time
to use elliptical galaxies as a test bench for stellar nucleosynthesis
studies.  At the same time we present several predictions for the
evolution of the N/Fe and N/O abundance in the gas that might be
tested on high-redshift proto-galaxies.

\section*{Acknowledgments} 
We wish to thank the referee for the insightful comments that improved
the quality of the paper. AP acknowledges financial support from the
Oxford University Astor Travel Grant and the hospitality of U.C. Santa
Cruz, where this project was begun, as well as partial support from
NSF grant AST-0649899 during the completion of the manuscript.  CC
acknowledges financial support from Swiss National Science Foundation
(SNF). F.M. acknowledge financial support by the Italian Space Agency
through contract ASI-INAF I/016/07/0. F.M. and C.C. acknowledge financial 
support from PRIN2007 (Italian Ministry of Research), Prot.2007JJC53X\_001.

\label{lastpage}

\end{document}